\documentclass[graybox]{svmult}

\usepackage{mathptmx}       
\usepackage{helvet}         
\usepackage{courier}        
\usepackage{type1cm}        
\usepackage{makeidx}         
\usepackage{graphicx}        
\usepackage{multicol}        
\usepackage[bottom]{footmisc}

\usepackage{amsmath}

\makeindex             

\newcommand{\DD}{\ensuremath{\mathsf{D}}}
\newcommand{\Id}{\ensuremath{\mathsf{I}}}

\begin{document}

\title*{Symbolic Computation of Conservation Laws, Generalized Symmetries,
and Recursion Operators for Nonlinear Differential-Difference 
Equations\thanks{This material is based upon work supported by 
the National Science Foundation (U.S.A.) under Grant No.\ CCF-0830783.}}
\titlerunning{Conservation Laws, Generalized Symmetries, and 
Recursion Operators of DDEs}
\author{\"Unal G\"okta\c{s} and Willy Hereman}
%
%
\institute{\"Unal G\"okta\c{s} \at Department of Computer Engineering, 
Turgut \"Ozal University,
Ke\c{c}i\"oren, Ankara 06010, Turkey \email{ugoktas@turgutozal.edu.tr}
\and Willy Hereman \at Department of Mathematical and Computer Sciences, 
Colorado School of Mines, Golden, Colorado 80401-1887, U.S.A. 
\email{whereman@mines.edu}}
%
%
\maketitle

\abstract*{Algorithms for the symbolic computation of polynomial 
conservation laws, generalized symmetries, and recursion operators for 
systems of nonlinear differential-difference equations (DDEs) are presented.
The algorithms can be used to test the complete integrability of nonlinear 
DDEs.
The ubiquitous Toda lattice illustrates the steps of the algorithms, 
which have been implemented in {\em Mathematica}.
The codes {\sc InvariantsSymmetries.m} and {\sc DDERecursionOperator.m} 
can aid researchers interested in properties of nonlinear DDEs.} 

\abstract{Algorithms for the symbolic computation of polynomial conservation 
laws, generalized symmetries, and recursion operators for systems of
nonlinear differential-difference equations (DDEs) are presented.
The algorithms
can be used to test the complete integrability of nonlinear DDEs.
The ubiquitous Toda lattice illustrates the steps of the algorithms, 
which have been implemented in {\em Mathematica}.
The codes {\sc InvariantsSymmetries.m} and {\sc DDERecursionOperator.m} 
can aid researchers interested in properties of nonlinear DDEs.}

\keywords{generalized symmetry\index{symmetry},
conservation law\index{conservation law},
recursion operator\index{recursion operator},
complete integrability\index{integrability},
differential-difference equation\index{differential-difference equation}.}
\section{Introduction}
A large number of physically important nonlinear models are completely
integrable, i.e., they can be linearized via an explicit transformation or 
can be solved with the Inverse Scattering Transform\index{IST}. 
Completely integrable continuous and discrete models arise in many branches 
of the applied sciences and engineering, including classical, quantum, 
and plasma physics, optics, electrical circuits, to name a few.
Mathematically, nonlinear models can be represented by ordinary 
and partial differential equations (ODEs and PDEs), differential-difference 
equations (DDEs), or ordinary and partial difference equations 
(O$\Delta$Es and P$\Delta$Es).
This paper deals with integrable nonlinear DDEs.

Completely integrable equations have nice analytic and geometric 
properties reflecting their rich mathematical structure. 
For instance, completely integrable PDEs and DDEs possess infinitely many 
conserved quantities and generalized (higher-order) symmetries of successive 
orders.
The existence of an infinite set of generalized symmetries can be 
established by explicitly constructing recursion operators which 
connect such symmetries. 
Finding generalized symmetries and recursion operators is a nontrivial task, 
in particular, if attempted by hand.
For example, in \cite{UG1998} and \cite{WHandUG1999} an algorithm
is presented to compute recursion operators for completely integrable PDEs, 
which was only recently implemented in {\it Mathematica} \cite{DBandWH2010}.

Based on our earlier work in \cite{UG1998}, 
\cite{UGandWHpd1998}, and \cite{UGandWHacm1999}, 
we present in this paper algorithms for the symbolic computation of 
conserved densities, generalized symmetries, and recursion operators of 
nonlinear systems of DDEs.
Such systems must be polynomial and of evolution type, i.e., the DDEs must 
be of first order in (continuous) time. 
The number of equations in the system, degree of nonlinearity, 
and order (shift levels) are arbitrary.
Furthermore, the current algorithms only cover {\it polynomial} densities, 
symmetries, and recursion operators.

We use the dilation (scaling) invariance of the system of DDEs to determine 
the {\it candidate} density, symmetry, or recursion operator. 
Indeed, these candidates are linear combinations with undetermined 
coefficients of scaling invariant terms. 
Upon substitution of the candidates into the corresponding defining equations,
one has to solve a linear system for the undetermined coefficients.
After doing so, the coefficients are substituted into the density, symmetry, 
or recursion operator. 
If so desired, the results can be tested one more time, in particular, by 
applying the recursion operators to generate the successive symmetries.

If the system of DDEs contains constant parameters, the eliminant of 
the linear system for the undetermined coefficients gives the necessary 
conditions for the parameters, so that the given DDEs admit the required 
density or symmetry. 
In analogy with the PDE case in \cite{UGandWH1997}, 
the algorithms can thus be used to classify DDEs with parameters according 
to their complete integrability as illustrated in 
\cite{UGandWHpd1998} and \cite{UGandWHacm1999}.

As shown in \cite{AF1980}, once the generalized symmetries are known, 
it is often possible to find the recursion operator by inspection.
If the recursion operator is {\em hereditary}, as defined in 
\cite{BFandWOandWW1987}, then the equation will possess infinitely many 
symmetries. 
If, in addition, the recursion operator is {\em factorizable} then the 
equation has infinitely many conserved quantities. 

Computer algebra systems can greatly help with the search for conservation 
laws, symmetries, and recursion operators. 
The algorithms in this paper have been implemented in {\em Mathematica}.
The computer codes (see \cite{WHwebsite2010}), can be used to test the 
complete integrability of systems of nonlinear DDEs, provided they are 
polynomial and of first order (or can be written in that form after 
a suitable transformation).

%
With {\sc InvariantsSymmetries.m}, in \cite{UG1998},
\cite{UGandWHpd1998}, and \cite{UGandWHacm1999}, 
G\"{o}kta\c{s} and Hereman computed polynomial conserved densities and 
generalized symmetries of many well-known systems of DDEs,
including various Volterra and Toda lattices as well as the Ablowitz-Ladik 
lattice (for additional results and references, see, e.g., 
\cite{Heremanetal2008}).
The existence of, say, a half dozen conserved densities or generalized 
symmetries is a {\it predictor} for complete integrability.
Finding a recursion operator then becomes within reach. 
%
An existence proof (showing that there are indeed infinitely many densities 
or generalized symmetries) must be done analytically, e.g., by explicitly 
constructing the recursion operator which allows one to generate the 
generalized symmetries order by order.
Numerous explicit examples have been reported in the literature but novices 
could start with the book by Olver \cite{PO1993} to learn about recursion 
operators for PDEs.  
To alleviate the burden of trying to find a recursion operator by trial 
and error, we present a new {\it Mathematica} program, 
{\sc DDERecursionOperator.m}, based on the algorithm in Section 5. 
Like {\sc InvariantsSymmetries.m}, after thorough testing,
{\sc DDERecursionOperator.m} will be available from \cite{WHwebsite2010}.

%
If one cannot find a sufficient large number of densities or symmetries 
(let alone, a recursion operator), then it is unlikely that the DDE system 
is completely integrable, at least in that coordinate representation. 
However, our software does not allow one to conclude that a DDE is {\it not}
completely integrable merely based on the fact that polynomial conserved 
densities and generalized symmetries could not be found. 
Polynomial DDEs that lack the latter may accidentally have non-polynomial 
densities or symmetries, or a complicated recursion operator, which is outside 
the scope of the algorithm described in Section 5.

Currently, our algorithm fails to find recursion operators for the 
Belov-Chaltikian lattices 
\cite{BelovChaltikian1993,SahadevanKhousalya2001,SahadevanKhousalya2003} 
and lattices due to Blaszak and Marciniak 
\cite{BlaszakMarciniak1994,SahadevanKhousalya2001,SahadevanKhousalya2003,WuGeng1996}. 
In the near future we plan to generalize the recursion operator algorithm 
so that it can cover a broader class of nonlinear DDEs. 


The paper is organized as follows. 
Basic definitions are given in Section 2.
In Section 3, we show the algorithm for conservation laws, 
using the Toda lattice as an example.  
Using the same example, Sections 4 and 5 cover the algorithms for 
generalized symmetries and recursion operators, respectively.
In Section 6, we draw some conclusions and briefly discuss future research.
%
\section{Key Definitions}
Consider a system of nonlinear DDEs of first order,
\begin{equation}
\label{DDEsystem}
{\dot{\bf u}}_n = {\bf F} ( {\bf u}_{n-\ell},
..., {\bf u}_{n-1}, {\bf u}_{n}, {\bf u}_{n+1}, ..., {\bf u}_{n+m}),
\end{equation}
where ${\bf u}_n$ and ${\bf F}$ are vector-valued functions with $N$ 
components. 
This paper only covers DDEs with {\it one} discrete variable, denoted by 
integer $n,$ which often corresponds to the discretization of a space 
variable. 
The dot stands for differentiation with respect to the continuous variable 
(often time $t).$ 
Each component of ${\bf F}$ is assumed to be a polynomial with constant 
coefficients.
If parameters are present in (\ref{DDEsystem}), they will be denoted by 
lower-case Greek letters.
${\bf F}$ depends on ${\bf u}_n$ and a finite number of forward and backward 
shifts of ${\bf u}_n.$ 
We denote by $\ell\, (m,$ respectively), the furthest negative 
(positive, respectively) shift of any variable in the system.
Restrictions are neither imposed on the degree of nonlinearity of ${\bf F},$
nor on the integers $l$ and $m,$ which measure the degree of non-locality 
in (\ref{DDEsystem}).
%
\subsection{Leading Example: The Toda Lattice}
One of the earliest and most famous examples of completely integrable DDEs 
is the Toda lattice, discussed in, for instance, \cite{MTbook1981}:
\begin{equation}
\label{todalatticeexponential} 
{\ddot{y}}_n = \exp{(y_{n-1} - y_n)} - \exp{(y_n - y_{n+1})},
\end{equation}
where $y_n$ is the displacement from equilibrium of the $n\/$th
particle with unit mass under an exponential decaying interaction
force between nearest neighbors. 
In new variables $(u_n, v_n),$ defined by $u_n = {\dot{y}}_n, 
v_n = \exp{(y_{n} - y_{n+1})},$ 
lattice (\ref{todalatticeexponential}) can be written in polynomial form
\begin{eqnarray}
\label{todalattice} 
{\dot{u}}_n &=& v_{n-1} - v_n, 
\nonumber \\
{\dot{v}}_n &=& v_n (u_n - u_{n+1}).
\end{eqnarray}
The Toda lattice (\ref{todalattice}) will be used to illustrate the various 
algorithms presented in subsequent sections of this paper.
%
\subsection{Dilation Invariance}
A DDE is {\em dilation invariant} if it is invariant under a dilation 
(scaling) symmetry.  
%
\subsubsection{Example}
Lattice (\ref{todalattice}) is invariant under scaling symmetry
\begin{equation}
\label{todascale}
(t, u_n, v_n) \rightarrow (\lambda^{-1} t, {\lambda}^1 u_n, {\lambda}^2 v_n).
\end{equation}
%
\subsection{Uniformity in Rank}
We define the {\em weight}, $w$, of a variable as the exponent of the 
scaling parameter $(\lambda)$ which multiplies that variable.
Since $\lambda$ can be selected at will, $t$ will always be replaced by 
$\frac{t}{\lambda}$ and, thus, $w(\frac{\rm{d}}{\rm{dt}}) = w({\rm D}_t) = 1.$ 

Weights of dependent variables are nonnegative, rational, and independent of 
$n.$
For example, $w(u_{n-3}) = \cdots = w(u_n) = \cdots = w(u_{n+2}).$ 

The {\em rank}, denoted by $R,$ of a monomial is defined as the total weight 
of the monomial.
An expression is {\em uniform in rank} if all of  its terms have the same rank.

Dilation symmetries, which are special Lie-point symmetries, are common 
to many DDEs. 
Polynomial DDEs that do not admit a dilation symmetry can be made scaling 
invariant by extending the set of dependent variables with auxiliary 
parameters with appropriate scales as discussed in \cite{UGandWHpd1998} 
and \cite{UGandWHacm1999}. 
%
\subsubsection{Example}
In view of (\ref{todascale}), we have $w(u_n)=1$, and $w(v_n)=2$ for 
the Toda lattice.
In the first equation of (\ref{todalattice}), all the monomials have
rank 2; in the second equation all the monomials have rank 3. 
Conversely, requiring uniformity in rank for each equation in 
(\ref{todalattice}) allows one to compute the weights of the dependent 
variables (and, thus, the scaling symmetry) with simple linear algebra. 
Balancing the weights of the various terms of each equation in 
(\ref{todalattice}) yields
\begin{eqnarray}
\label{todaweightequations}
w(u_n) + 1 &=& w(v_n), \nonumber \\
w(v_n) + 1 &=& w(u_n) + w(v_n). 
\end{eqnarray}
Hence,
\begin{equation}
\label{todaweights}
w(u_n) = 1, \quad w(v_n) = 2, 
\end{equation} 
which confirms (\ref{todascale}).
%
\subsection{Up-Shift and Down-Shift Operator}
We define the shift operator $\DD$ by $\DD{\bf u}_n = {\bf u}_{n+1}.$ 
The operator $\DD$ is often called the {\em up-shift operator} or forward- 
or right-shift operator. 
The inverse, $\DD^{-1},$ is the {\em down-shift operator} or backward- 
or left-shift operator, $\DD^{-1} {\bf u}_n = {\bf u}_{n-1}.$ 
Shift operators apply to functions by their action on the arguments 
of the functions. 
For example,
\begin{eqnarray}
\label{Dfunction}
&& \DD {\bf F}({\bf u}_{n-\ell}, \cdots, 
{\bf u}_{n-1}, {\bf u}_{n}, {\bf u}_{n+1}, \cdots,{\bf u}_{n+m})
\nonumber \\
&& \quad = {\bf F} (\DD {\bf u}_{n-\ell}, \cdots, 
\DD {\bf u}_{n-1}, \DD {\bf u}_{n}, \DD {\bf u}_{n+1}, \dots, 
\DD {\bf u}_{n+m})
\nonumber \\
&& \quad = {\bf F} ({\bf u}_{n-\ell+1}, \dots, {\bf u}_{n}, {\bf u}_{n+1}, 
{\bf u}_{n+2}, \cdots, {\bf u}_{n+m+1}).
\end{eqnarray}
%
\subsection{Conservation Law}
A {\em conservation law} of (\ref{DDEsystem}),
\begin{equation}
\label{conslawdde}
\DD_t \, \rho + \Delta \, J = 0,
\end{equation}
connects a {\em conserved density} $\rho$ to an {\em associated flux} $J,$ 
where both are scalar functions depending on ${\bf u}_n$ and its shifts. 
In (\ref{conslawdde}), which {\it must holds} on solutions of 
(\ref{DDEsystem}), 
$\DD_t$ is the total derivative with respect to time,
$\Delta = \DD - \Id$ is the {\it forward difference operator}, 
and $\Id$ is the identity operator.
For readability (in particular, in the examples), the components of 
${\bf u}_n$ will be denoted by $u_n, v_n, w_n, $ etc.\ 
In what follows we consider only autonomous functions, 
i.e., ${\bf F}, \rho,$ and $J$ do not explicitly depend on $t$ and $n.$

A density is {\em trivial} if there exists a function $\psi$ so that
$\rho = \Delta \psi.$
We say that two densities, $\rho^{(1)}$ and $\rho^{(2)},$ 
are {\em equivalent} if and only if $\rho^{(1)} + k \rho^{(2)} = \Delta \psi,$
for some $\psi$ and some non-zero scalar $k.$
It is paramount that the density is free of equivalent terms for if such 
terms were present, they could be moved into the flux $J.$

Compositions of $\DD$ or $\DD^{-1}$ define an {\em equivalence relation\/} 
$(\equiv)$ on monomial terms.
Simply stated, all shifted terms are {\em equivalent},
e.g., $u_{n-1} v_{n+1} \equiv u_n v_{n+2} 
\equiv u_{n+2} v_{n+4} \equiv u_{n-3} v_{n-1}$
since
\begin{eqnarray}
\label{equivalenceexample}
u_{n-1} v_{n+1} &=& u_n v_{n+2} - \Delta  ( u_{n-1} v_{n+1}) 
\nonumber\\
&=& u_{n+2} v_{n+4} - \Delta ( u_{n+1} v_{n+3} 
    + u_n v_{n+2} + u_{n-1} v_{n+1} ) 
\nonumber\\
&=& u_{n-3} v_{n-1} + \Delta  ( u_{n-2} v_n + u_{n-3} v_{n-1}).
\end{eqnarray}
This equivalence relation also holds for any function of the dependent
variables, but for the construction of conserved densities we will apply it 
only to monomial terms $(t_i)$ in the {\it same} density, 
thereby achieving high computational efficiency.
%
In the algorithm used in Section 3, we will use the following 
{\em equivalence criterion}: 
two monomial terms, $t_1$ and $t_2$, are equivalent, 
$t_1 \equiv t_2,$ if and only if $t_1 = \DD^r \, t_2$ 
for some integer $r.$ 
If $t_1 \equiv t_2$ then $t_1 = t_2 + \Delta J$ 
for some $J$ dependent on ${\bf u}_n$ and its shifts. 
For example, $u_{n-2} u_n \equiv u_{n-1} u_{n+1}$ because
$u_{n-2} u_n = \DD^{-1} u_{n-1} u_{n+1}.$ 
Hence, $u_{n-2} u_n = u_{n-1} u_{n+1} + [- u_{n-1} u_{n+1} + u_{n-2} u_n ] 
= u_{n-1} u_{n+1} + \Delta J$ with $J = -u_{n-2} u_n .$

For efficiency, we need a criterion to choose a unique representative
from each equivalence class. 
There are a number of ways to do this.
We define the {\em canonical} representative as that member that has
(i) no negative shifts and (ii) a non-trivial dependence on the {\em local} 
(that is, zero-shifted) variable. 
For example, 
$u_n u_{n+2}$ is the canonical representative of the class 
\[ 
\{ \cdots, u_{n-2} u_n, u_{n-1} u_{n+1}, u_n u_{n+2}, u_{n+1} u_{n+3}, 
\cdots \}. 
\] 
In the case of, e.g., two variables $(u_n$ and $v_{n})$, $u_{n+2} v_n$ is the 
canonical representative of the class
\[ 
\{ \cdots, u_{n-1} v_{n-3}, u_n v_{n-2}, u_{n+1} v_{n-1}, u_{n+2} v_n, 
u_{n+3} v_{n+1}, \cdots \}. 
\]
Alternatively, one could choose a variable ordering and then choose the 
member that depends on the zero-shifted variable of lowest
lexicographical order. 
The code in \cite{WHwebsite2010} uses lexicographical ordering of the 
variables, i.e., $u_n \prec v_n \prec w_n,$ etc. 
Thus, $u_n v_{n-2}$ (instead of $u_{n+2} v_n)$ is chosen as the
canonical representative of $\{ \cdots, u_{n-1} v_{n-3}, u_n v_{n-2}, 
u_{n+1} v_{n-1}, u_{n+2} v_n, u_{n+3} v_{n+1}, \cdots \}.$

It was shown in \cite{HickmanJNMP2008} that if $\rho$ is a density then 
$\DD^k \rho$ is also a density. 
Hence, using an appropriate ``up-shift" all negative shifts in a density 
can be removed. 
Without loss of generality, we thus assume that a density that depends on 
$q$ shifts has {\em canonical} form 
$\rho({\bf u}_n, {\bf u}_{n+1}, \cdots, {\bf u}_{n+q}).$ 
%
\subsubsection{Example}
Lattice (\ref{todalattice}) has infinitely many conservation laws
(see, e.g., \cite{MHphysrev1974}). 
Here we list the densities of rank $R \leq 4:$
\begin{eqnarray}
\label{todarho1} 
\rho^{(1)} &=& u_n, 
\\
\label{todarho2} 
\rho^{(2)} &=& \tfrac{1}{2}{u_n^2} + v_n,
\\
\label{todarho3} 
\rho^{(3)} &=& \tfrac{1}{3}{u_n^3} + u_n (v_{n-1} + v_n),
\\
\label{todarho4} 
\rho^{(4)} &=& \tfrac{1}{4}{u_n^4} + {u_n^2} (v_{n-1} + v_n) + u_n u_{n+1} v_n 
+\tfrac{1}{2} {v_n^2} + v_n v_{n+1}.
\end{eqnarray}
The first two density-flux pairs are easily computed by hand, and so is
\begin{equation}
\label{condenstoda0}
\rho_n^{(0)} = \ln (v_n), 
\end{equation}
which is the only non-polynomial density (of rank $0).$
%
\subsection{Generalized Symmetry}
A vector function ${\bf G}({\bf u}_n)$ is called a {\em generalized symmetry} 
of (\ref{DDEsystem}) if the infinitesimal transformation 
${\bf u}_n \rightarrow {\bf u}_n + \epsilon {\bf G}$ 
leaves (\ref{DDEsystem}) invariant up to order $\epsilon.$ 
As shown by \cite{PO1993}, ${\bf G}$ must then satisfy
\begin{equation}
\label{ddesymmetry}
{\rm D}_{t}{\bf G} = {\bf F}'({\bf u}_n)[{\bf G}]
\end{equation}
on solutions of (\ref{DDEsystem}), where ${\bf F}'({\bf u}_n)[{\bf G}]$ 
is the Fr\'echet derivative of ${\bf F}$ in the direction of ${\bf G}.$ 

For the scalar case $(N=1)$, the Fr\'echet derivative is
\begin{equation}
\label{ddefrechetscalar}
F'(u_n)[G] = 
\frac{\partial}{\partial{\epsilon}} F(u_n +\epsilon G) {|_{\epsilon = 0}}
= \sum_{k} \frac{\partial F}{\partial u_{n+k}} {\rm D}^{k} G,
\end{equation}
which, in turn, defines the Fr\'echet derivative operator
\begin{equation}
\label{ddefrechetscalaroperator}
F'(u_n) = \sum_{k} \frac{\partial F}{\partial u_{n+k}} {\rm D}^{k}.
\end{equation}
In the vector case with, say, components $u_n$ and $v_n,$ 
the Fr\'echet derivative operator is a matrix operator:
\begin{equation}
\label{ddefrechetvectoroperator}
{\bf F}'({\bf u}_n)  
= \left(\; \begin{array}{cc} 
\sum_{k} \frac{\partial F_1}{\partial u_{n+k}} {\rm D}^{k} \;
& \;\; \sum_{k} \frac{\partial F_1}{\partial v_{n+k}} {\rm D}^{k} 
\\ 
& \\
\sum_{k} \frac{\partial F_2}{\partial u_{n+k}} {\rm D}^{k} \;
& \;\; \sum_{k} \frac{\partial F_2}{\partial v_{n+k}} {\rm D}^{k}
\end{array} \; \right).
\end{equation}
Applied to ${\bf G} = (G_1 \;\; G_2)^{\rm T},$ where ${\rm T}$ is transpose, 
one obtains
\begin{equation}
\label{ddefrechetcomponent}
{F_i}'({\bf u}_n)[{\bf G}] 
= \sum_{k} \frac{\partial F_i}{\partial u_{n+k}} {\rm D}^{k} G_1
+ \sum_{k} \frac{\partial F_i}{\partial v_{n+k}} {\rm D}^{k} G_2, 
\end{equation}
with $i = 1, 2.$
In (\ref{ddefrechetscalar}) and (\ref{ddefrechetcomponent}) 
summation is over all positive and negative shifts (including $k=0).$
The generalization of (\ref{ddefrechetvectoroperator}) to a 
$N-$component system is straightforward.
%
\subsubsection{Example}
As computed in \cite{WHandUGandMCandAM98}, the first two non-trivial 
symmetries of (\ref{todalattice}) are
\begin{eqnarray}
\label{todasymm1}
{\bf G}^{(1)} 
&=& \left( \begin{array}{c} 
v_{n}-v_{n-1} \\ 
\nonumber \\
v_{n}(u_{n+1}-u_{n}) 
\end{array} \right), \\ 
\nonumber \\
\label{todasymm2}
{\bf G}^{(2)} 
&=& \left( \begin{array}{c} 
v_{n}(u_{n} + u_{n+1}) - v_{n-1}(u_{n-1}+u_{n}) \\
\nonumber \\
v_{n}(u_{n+1}^{2} - u_{n}^{2} + v_{n+1} - v_{n-1})
\end{array} \right).
\end{eqnarray}
%
\subsection{Recursion Operator}
A {\em recursion operator} ${\mathcal R}$ connects symmetries 
\begin{equation}
\label{symmetrylink}
{\bf G}^{(j+s)} = {\mathcal R} \, {\bf G}^{(j)}, 
\end{equation}
where $ j=1, 2, \cdots ,$ and $s$ is the gap length. 
The symmetries are linked consecutively if $s=1.$ 
This happens in most (but not all) cases.
For $N\/$-component systems, ${\mathcal R}$ is an 
$N \times N$ matrix operator.

With reference to \cite{PO1993} and \cite{JW1998}, the defining equation 
for ${\mathcal R}$ is
\begin{eqnarray}
\label{definingrecursion}
&& {\rm D}_t {\mathcal R} + [{\mathcal R}, {\bf F}'({\bf u}_n)] 
\nonumber \\
&& \quad = \frac{\partial {\mathcal R}}{\partial t}
+ {\mathcal R}' [{\bf F}] + {\mathcal R} \circ {\bf F}'({\bf u}_n) 
- {\bf F}'({\bf u}_n)\circ {\mathcal R} = 0, 
\end{eqnarray}
%
where $[\; , \; ]$ denotes the commutator and $\circ$ the composition of 
operators. 
The operator ${\bf F}'({\bf u}_n)$ was defined in 
(\ref{ddefrechetvectoroperator}).
${{\mathcal R}}' [{\bf F}]$ is the Fr\'echet derivative of 
${\mathcal R}$ in the direction of ${\bf F}.$ 
For the scalar case, the operator ${\mathcal R}$ is often of the form

\begin{equation}
\label{recursionoperatorscalar}
{\mathcal R} = 
U(u_n) \; 
{\mathcal O} \left( 
({\rm D} - {\rm I})^{-1}, {\rm D}^{-1}, {\rm I}, {\rm D} \right) 
\; V(u_n),
\end{equation}
and then
\begin{equation}
\label{frechetofrscalar}
{\mathcal R}' [F] = 
\sum_{k}({\rm D}^k F)\frac{\partial U}{\partial u_{n+k}}{\mathcal O} \, V
+\sum_{k}U {\mathcal O} ({\rm D}^k F)\frac{\partial V}{\partial u_{n+k}}.
\end{equation}
For the vector case, the elements of the $N \times N$ operator matrix 
${\mathcal R}$ are often of the form 
\begin{equation}
\label{recursionoperatorvector}
{\mathcal R}_{ij} = 
U_{ij}({\bf u}_n) \, 
{\mathcal O}_{ij}\left( 
({\rm D} - {\rm I})^{-1}, {\rm D}^{-1}, {\rm I}, {\rm D} \right) 
\, V_{ij}({\bf u}_n).
\end{equation}
Hence, for the 2-component case 
\begin{eqnarray}
\label{frechetofrvector}
{\mathcal R}'[{\bf F}]_{ij} &=&
\sum_{k} \, ({\rm D}^k F_1) \, 
\frac{\partial U_{ij}}{\partial u_{n+k}} \, {\mathcal O}_{ij} \, 
V_{ij} 
+ \sum_{k} \, ({\rm D}^k F_2) \, 
\frac{\partial U_{ij}}{\partial v_{n+k}} \, 
{\mathcal O}_{ij} \, V_{ij} 
\nonumber \\
&& + \sum_{k} \, U_{ij} \, {\mathcal O}_{ij} \, 
({\rm D}^k F_1) \,\frac{\partial V_{ij}}{\partial u_{n+k}} 
+ \sum_{k} \, U_{ij} \, {\mathcal O}_{ij} \, 
({\rm D}^k F_2) \,\frac{\partial V_{ij}}{\partial v_{n+k}}.
\end{eqnarray}
%
\subsubsection{Example}
The recursion operator of (\ref{todalattice}) is
\begin{equation}
\label{recursiontoda}
{\mathcal R} 
= \left(\, \begin{array}{cc} u_{n}{\rm I} \;
& \;\; {\rm D}^{-1}+{\rm I} 
+ (v_{n} - v_{n-1})({\rm D} - {\rm I})^{-1}\,\frac{1}{v_{n}}\, {\rm I} 
\\ 
& 
\\
v_n {\rm I}+v_n{\rm D} \;
& 
\;\; u_{n+1}{\rm I} + v_{n}(u_{n+1} - u_{n})({\rm D} - {\rm I})^{-1} 
\frac{1}{v_{n}} {\rm I}
\end{array} \; \right).
\end{equation}
It is straightforward to verify that ${\mathcal R}\, G^{(1)} = G^{(2)}$ 
with $G^{(1)}$ in (\ref{todasymm1}) and $G^{(2)}$ in (\ref{todasymm2}). 
%
\section{Algorithm for Conservation Laws}
As an example, we will compute the density $\rho^{(3)}$ (of rank $R = 3)$ 
given in (\ref{todarho3}).
%
\subsection{Construct the Form of the Density}
Start from ${\cal V} = \{ u_n, v_n \},$ the set of dependent variables with 
weights. 
List all monomials in $u$ and $v$ of rank $R = 3$ or less: 
${\cal M} = \{ u_n^3, u_n^2, u_n v_n, u_n, v_n \}.$ 
Next, for each monomial in ${\cal M}$, introduce the correct number of 
$t$-derivatives so that each term has rank $3.$
Using (\ref{todalattice}), compute
\begin{eqnarray}
\label{todaweightadjust} 
&& \frac{{\rm d}^0 u_n^3}{ {\rm dt}^0} = u_n^3, 
\nonumber \\
&& \frac{{\rm d}^0 u_n v_n}{{\rm dt}^0} = u_n v_n, 
\nonumber \\
&& \frac{{\rm d} u_n^2}{{\rm dt}} 
= 2 u_n {\dot u}_n = 2 u_n v_{n-1} - 2 u_n v_n, 
\\
&& \frac{{\rm d} v_n}{{\rm dt}} = 
{\dot v}_n =  u_n v_n - u_{n+1} v_n, 
\nonumber \\
&& \frac{{\rm d}^2 u_n}{{\rm dt}^2} = 
\frac{{\rm d}{\dot u}_n}{{\rm dt}}
   = \frac{{\rm d} (v_{n-1} - v_n)}{{\rm dt}} \nonumber \\
&& \quad\quad\quad\quad\quad\;
= u_{n-1} v_{n-1} - u_n v_{n-1} - u_n v_n + u_{n+1} v_n. 
\nonumber
\end{eqnarray}
Gather the terms in the right hand sides in (\ref{todaweightadjust})
to get 
\[{\cal S} = \{ u_n^3, u_n v_{n-1}, u_n v_n, u_{n-1} v_{n-1}, u_{n+1} v_n \}.\]

Identify members belonging to the same equivalence classes and
replace them by their canonical representatives.
For example, $u_n v_{n-1} \equiv u_{n+1} v_n.$ 
Adhering to lexicographical ordering, use $u_n v_{n-1}$ instead of 
$u_{n+1} v_n.$ 
Doing so, replace ${\cal S}$ by 
${\cal T} = \{ u_n^3, u_n v_{n-1}, u_n v_n \},$ 
which has the building blocks of the density. 
Linearly combine the monomials in ${\cal T}$ with undetermined coefficients 
$c_i$ to get the candidate density of rank $3:$
\begin{equation}
\label{formrho3toda} 
\rho = c_1 \, u_n^3 + c_2 \, u_n v_{n-1} + c_3 \, u_n v_n.
\end{equation}
%
\subsection{Compute the Undetermined Coefficients $c_i$}
Compute $\DD_t \rho$ and use (\ref{todalattice}) to eliminate 
${\dot u}_n$ and ${\dot v}_n$ and their shifts. 
Next, introduce the main representatives to get
\begin{eqnarray}
\label{Etodalattice} E &=&
  (3 c_1 - c_2 ) u_n^2 v_{n-1} + (c_3 - 3 c_1 ) u_n^2 v_n 
  + (c_3 - c_2) v_{n} v_{n+1}
\nonumber \\
&& +\, (c_2 - c_3) u_{n} u_{n+1} v_{n} + (c_2 - c_3) v_{n}^2 
+ \Delta J, 
\end{eqnarray}
with
\begin{equation}
\label{formJ3toda}
J = (c_3 - c_2) v_{n-1} v_n + c_2 u_{n-1} u_n v_{n-1} + c_2 v_{n-1}^2.
\end{equation}
%
Set $E - \Delta J \equiv 0$ to get the linear system
\begin{equation}
\label{systemtodacanonicalform} 3 c_1 - c_2 = 0, \quad c_3 - 3 c_1 = 0, 
\quad c_2 - c_3 = 0.
\end{equation}
Select $c_1 = \tfrac{1}{3}$ and substitute the solution 
$c_1 =\tfrac{1}{3}, c_2 = c_3 = 1,$ into (\ref{formrho3toda}) and 
(\ref{formJ3toda}) to obtain $\rho^{(3)}$ in (\ref{todarho3}) with
matching flux $J^{(3)} = u_{n-1} u_n v_{n-1} + v_{n-1}^2.$
%
\section{Algorithm for Symmetries}
As an example, we will now compute the symmetry 
${\bf G}^{(2)} = ( G_1^{(2)} \;\;\; G_2^{(2)} )^{\rm T}$ with 
${\rm rank}\, {\bf G} = (3 \;\;\; 4)^{\rm T}$ given in (\ref{todasymm2}).
%
\subsection{Construct the Form of the Symmetry}
Listing all monomials in $u_n$ and $v_n$ of ranks $3$ and 4, or less: 
\begin{eqnarray}
{\cal L}_1 &=& \{ u_n^3, u_n^2, u_n v_n, u_n, v_n \}, 
\nonumber \\
{\cal L}_2 &=& 
\{ u_n^4, u_n^3, u_n^2 v_n, u_n^2, u_n v_n, u_n, v_n^2, v_n \}. 
\nonumber 
\end{eqnarray}
Next, for each monomial in ${\cal L}_1$ and ${\cal L}_2$, 
introduce the necessary $t$-derivatives so that each term exactly has 
ranks $3$ and $4$, respectively. 
At the same time, use (\ref{todalattice}) to remove all $t-$derivatives.
Doing so, based on ${\cal L}_1,$ 
\begin{eqnarray}
\label{todaweightadjust2}
&&{{\rm{d}}^0 \over {\rm{dt}}^0} ( u_n^3 )= u_n^3 , \nonumber \\
&& {{\rm{d}}^0 \over {\rm{dt}}^0} ( u_n v_n ) = u_n v_n , \nonumber \\
&& {{\rm{d}} \over {\rm{dt}}}( u_n^2 ) 
= 2 u_n {\dot{u}}_n = 2 u_n v_{n-1} - 2 u_n v_n , \\
&& {{\rm{d}} \over {\rm{dt}}} ( v_n )
= {\dot{v}}_n =  u_n v_n -  u_{n+1} v_n, \nonumber \\
&& {{\rm{d}}^2 \over {\rm{dt}}^2} ( u_n )
   = {{\rm{d}} \over {\rm{dt}}} ( {\dot{u}}_n )
   = {{\rm{d}} \over {\rm{dt}}} ( v_{n-1} - v_n ) \nonumber \\
&& \quad\quad\quad\quad\quad\quad\quad\;\;
= u_{n-1} v_{n-1} - u_{n} v_{n-1} - u_n v_n + u_{n+1} v_n .
\nonumber
\end{eqnarray}
Put the terms from the right hand sides of (\ref{todaweightadjust2}) 
into a set: 
\[{\cal W}_1 = 
\{ u_n^3, u_{n-1} v_{n-1} , u_n v_{n-1} , u_n v_n , u_{n+1} v_n \}.
\]
\vskip 1pt
\noindent
Similarly, based on the monomials in ${\cal L}_2,$ construct
\begin{eqnarray*}
{\cal W}_2 & = &
\{ u_n^4, u_{n-1}^2 v_{n-1}, u_{n-1} u_n v_{n-1}, 
u_n^2 v_{n-1}, v_{n-2} v_{n-1}, v_{n-1}^2,  \\
&&\quad u_n^2 v_n, u_n u_{n+1} v_n, u_{n+1}^2 v_n, v_{n-1} v_n, 
v_n^2, v_n v_{n+1} \}.
\end{eqnarray*}
Linearly combine the monomials in ${\cal W}_1$ and ${\cal W}_2$ 
with undetermined coefficients $c_i$ to get the form of the components of 
the candidate symmetry:
\begin{eqnarray}
\label{formsym3toda}
G_1^{(2)}
&=& c_1 \, u_n^3 + c_2 \, u_{n-1} v_{n-1} + c_3 \, u_n v_{n-1} 
+ c_4 \, u_n v_n + c_5 \, u_{n+1} v_n, 
\nonumber \\
&& \\
G_2^{(2)} &=& c_6 \, u_n^4 + c_7 \, u_{n-1}^2 v_{n-1} + 
c_8 \, u_{n-1} u_n v_{n-1} + c_9 \, u_n^2 v_{n-1}  
\nonumber \\ 
&& +\, c_{10} \, v_{n-2} v_{n-1}+ c_{11} \, v_{n-1}^2 + c_{12} \, u_n^2 v_n 
+ c_{13} \, u_n u_{n+1} v_n 
\nonumber \\
&& +\, c_{14} \, u_{n+1}^2 v_n +c_{15} \, v_{n-1} v_n + c_{16} \, v_n^2 
+ c_{17} \, v_n v_{n+1} . 
\nonumber
\end{eqnarray} 
%
\subsection{Compute the Undetermined Coefficients $c_i$}
To determine the coefficients $c_i,$ require that (\ref{ddesymmetry}) 
holds on any solution of (\ref{DDEsystem}).
Compute ${\rm D}_t {\bf G}$ and use (\ref{DDEsystem}) to remove all 
$ {\dot {\bf u}}_{n-1}, {\dot {\bf u}}_n, {\dot {\bf u}}_{n+1},$ etc. 
Compute the Fr\'echet derivative (\ref{ddefrechetcomponent}) and, 
in view of (\ref{ddesymmetry}), equate the resulting expressions. 
Treat as independent all the monomials in ${\bf u}_n $ and their shifts, 
to obtain the linear system that determines the coefficients $c_i.$

Apply the strategy to (\ref{todalattice}) with (\ref{formsym3toda}), to get
\begin{eqnarray*}
\label{sym3todaresult}
c_1 = c_6 = c_7 = c_8 = c_9 = c_{10} = c_{11} = c_{13} = c_{16} = 0, \\
-c_2 = -c_3 = c_4 = c_5 = -c_{12} = c_{14} = -c_{15} = c_{17}.
\end{eqnarray*}
Set $c_{17} = 1$ and substitute (\ref{sym3todaresult}) into 
(\ref{formsym3toda}) to get ${\bf G}^{(2)} = 
(G_1^{(2)} \;\;\; G_2^{(2)})^{\rm T},$ as given in (\ref{todasymm2}).

To show how our algorithm filters out completely integrable cases 
among parameterized systems of DDEs, consider  
\begin{eqnarray}
\label{partodalatt}
{\dot{u}}_n &=& \alpha \; v_{n-1} - v_n, 
\nonumber \\
{\dot{v}}_n &=& v_n \; (\beta \; u_n - u_{n+1}),
\end{eqnarray}
where $\alpha$ and $\beta$ are {\em nonzero} constant parameters.
\cite{ARandBGandKT1992} have shown that (\ref{partodalatt}) is completely 
integrable if and only if $\alpha = \beta = 1.$

Using our algorithm, one can easily compute the {\em compatibility conditions}
for $\alpha$ and $\beta$ so that (\ref{partodalatt}) admits a polynomial
symmetry, say, of rank $(3,4)$.
The steps are as above, however, the linear system for the $c_i$ is 
parameterized by $\alpha$ and $\beta$ and must be analyzed carefully 
(with, e.g., Gr\"obner basis methods).
This analysis leads to the condition $ \alpha = \beta = 1.$
Details are given in \cite{UGandWHpd1998} and \cite{UGandWHacm1999}. 
%
\section{Algorithm for Recursion Operators}
We will now construct the recursion operator (\ref{recursiontoda}) for 
(\ref{todalattice}). 
In this case all the terms in (\ref{definingrecursion}) are $2 \times 2$ 
matrix operators.
%
\subsection{Determine the Rank of the Recursion Operator}
The difference in the ranks of symmetries is used to compute the rank of the 
elements of the recursion operator.
Use (\ref{todaweights}), (\ref{todasymm1}) and (\ref{todasymm2}) to compute
\begin{equation}
\label{rankrtoda}
{\rm rank} \, {\bf G}^{(1)}
= \left( \begin{array}{c}
2 \\ 
3
\end{array} \right), 
\quad
{\rm rank}\, {\bf G}^{(2)}
= \left( \begin{array}{c}
3 \\ 
4
\end{array} \right).
\end{equation}
Assume that ${\mathcal R} \, {\bf G}^{(1)} = {\bf G}^{(2)}$ and use 
the formula
\begin{equation}
\label{recursionrankrules}
{\rm rank}\, {\mathcal R}_{ij}
= {\rm rank}\, G^{(k+1)}_{i} - {\rm rank}\, G^{(k)}_{j}, 
\end{equation}
to compute a rank matrix associated to the operator ${\mathcal R}:$
\begin{equation}
\label{todarankmatrix}
{\rm rank} \, {\mathcal R} 
= \left( \begin{array}{cc}
1 \; & \; 0
\\
2 \; & \; 1
\end{array} \right).
\end{equation}
%
\subsection{Determine the Form of the Recursion Operator}
We assume that ${\mathcal R} = {\mathcal R}_{0}+{\mathcal R}_{1},$ 
where ${\mathcal R}_{0}$ is a sum of terms involving
${\rm D}^{-1}, {\rm I},$ and ${\rm D}.$ 
(The form of ${\mathcal R}_{1}$ will be discussed below.)
The coefficients of these terms are admissible power combinations of 
$u_n, u_{n+1}, v_n,$ and $v_{n-1}$ 
(which come from the terms on the right hand sides of (\ref{todalattice})),
so that all the terms have the correct rank. 
The maximum up-shift and down-shift operator that should be included can be 
determined by comparing two consecutive symmetries.
Indeed, if the maximum up-shift in the first symmetry is $u_{n+p}$ 
and the maximum up-shift in the next symmetry is $u_{n+p+r},$ 
then the associated piece that goes into ${\mathcal R}_{0}$ must have
${\rm D}, {\rm D}^2, \ldots,{\rm D}^r.$ 
The same argument determines the minimum down-shift operator to be included. 
For (\ref{todalattice}), get
\begin{eqnarray}
\label{r0candidatetoda}
{\mathcal R}_{0} 
&=& \left( \begin{array}{cc}
({\mathcal R}_{0})_{11}\;\; &\;\; ({\mathcal R}_{0})_{12} \\
({\mathcal R}_{0})_{21}\;\; &\;\; ({\mathcal R}_{0})_{22}
\end{array} \right), 
\end{eqnarray}
with
\begin{eqnarray}
\label{recursionoperatorelements}
({\mathcal R}_{0})_{11} 
&=& (c_{1}u_{n} + c_{2}u_{n+1})\, {\rm I},
\nonumber \\
({\mathcal R}_{0})_{12} 
&=& c_{3}{\rm D}^{-1} + c_{4}{\rm I},
\nonumber \\
({\mathcal R}_{0})_{21} 
&=& (c_{5}u_{n}^{2} + c_{6}u_{n}u_{n+1} + c_{7}u_{n+1}^{2}
+ c_{8}v_{n-1} + c_{9}v_{n})\, {\rm I}
\\ 
&& +\, (c_{10}u_{n}^{2} + c_{11}u_{n}u_{n+1} + c_{12}u_{n+1}^{2}
+ c_{13}v_{n-1} + c_{14} v_{n})\, {\rm D},
\nonumber \\
({\mathcal R}_{0})_{22} &=& (c_{15}u_{n} + c_{16}u_{n+1})\, {\rm I}.
\nonumber 
\end{eqnarray}
As shown for the continuous case \cite{WHandUG1999}, 
${\mathcal R}_{1}$ is a linear combination (with undetermined coefficients 
${\tilde c}_{jk})$ of all suitable products of symmetries and covariants, 
i.e., Fr\'echet derivatives of densities, 
sandwiching $({\rm D}-{\rm I})^{-1}.$
Hence,
\begin{equation}
\label{operatorr1}
\sum_j \sum_k {\tilde c}_{jk} {\bf G}^{(j)} ({\rm D}-{\rm I})^{-1} 
\otimes \rho_n^{(k)'}, 
\end{equation}
where $\otimes$ denotes the matrix outer product, defined as
\begin{eqnarray}
\label{outerproduct}
&&
\left( \begin{array}{c}
G_1^{(j)} \\ 
\\
G_2^{(j)} \end{array} \right) 
({\rm D} - {\rm I})^{-1} \otimes
\left( \rho_{n,1}^{(k)\prime}\;\;\; \rho_{n,2}^{(k)\prime} \right)
= \nonumber \\
&&\quad\quad\quad \left(\; \begin{array}{cc}
G_1^{(j)} ({\rm D}-{\rm I})^{-1}\rho_{n,1}^{(k)\prime} \;\; & 
\;\; G_1^{(j)} ({\rm D}-{\rm I})^{-1}\rho_{n,2}^{(k)\prime}
\\ 
& 
\\
G_2^{(j)} ({\rm D} - {\rm I})^{-1} \rho_{n,1}^{(k)\prime} \;\; & 
\;\; G_2^{(j)} ({\rm D} - {\rm I})^{-1} \rho_{n,2}^{(k)\prime}
\end{array} \; \right).
\end{eqnarray}
Only the pair $({\bf G}^{(1)},\rho_n^{(0)\prime})$ can be used, 
otherwise the ranks in (\ref{todarankmatrix}) would be exceeded. 
Use (\ref{condenstoda0}) and (\ref{ddefrechetcomponent}), to compute 
\begin{equation}
\rho_n^{(0)\prime} = 
\left(\; \begin{array}{cc} {\rm 0} \;\; &\;
\frac{1}{v_n} {\rm I} \end{array} \; \right), 
\end{equation}
From (\ref{operatorr1}), after renaming ${\tilde c}_{10}$ to $c_{17},$ obtain
\begin{equation}
\label{r1candidatetoda}
{\mathcal R}_{1} 
= \left(\; \begin{array}{cc}
{\rm 0} \;\; & 
\;\; c_{17}\, 
(v_{n-1} -v_{n})({\rm D} -{\rm I})^{-1} \, \frac{1}{v_{n}}\, {\rm I}
\\
& 
\\ 
{\rm 0} \;\; & 
\;\; c_{17}\, v_{n}(u_{n} - u_{n+1})({\rm D}- {\rm I})^{-1} \, 
\frac{1}{v_{n}} \, {\rm I}
\end{array} \; \right).
\end{equation}
Add (\ref{r0candidatetoda}) and (\ref{r1candidatetoda}), to get
\begin{eqnarray}
\label{rcandidatetoda}
{\mathcal R} = {\mathcal R}_{0} + {\mathcal R}_{1}
&=& \left( \begin{array}{cc}
{\mathcal R}_{11}\;\; &\;\; {\mathcal R}_{12} \\
{\mathcal R}_{21}\;\; &\;\; {\mathcal R}_{22}
\end{array} \right), 
\end{eqnarray}
with
\begin{eqnarray}
\label{operatorElements2}
{\mathcal R}_{11} 
&=& (c_{1}u_{n} + c_{2}u_{n+1})\, {\rm I}, 
\nonumber \\
{\mathcal R}_{12} 
&=& c_{3}{\rm D}^{-1} + c_{4}{\rm I}
    + c_{17}(v_{n-1}-v_{n})({\rm D}-{\rm I})^{-1}\, \frac{1}{v_{n}} \,{\rm I},
\nonumber \\
{\mathcal R}_{21} 
&=& (c_{5}u_{n}^{2} + c_{6}u_{n}u_{n+1} + c_{7}u_{n+1}^{2}
    + c_{8}v_{n-1} + c_{9}v_{n})\, {\rm I}
\\ 
&&  +\, (c_{10}u_{n}^{2} + c_{11}u_{n}u_{n+1} + c_{12}u_{n+1}^{2}
    + c_{13}v_{n-1} + c_{14}v_{n}) {\rm D},
\nonumber \\
{\mathcal R}_{22} 
&=& (c_{15}u_{n} + c_{16}u_{n+1}){\rm I}
    + c_{17}v_{n}(u_{n}-u_{n+1})({\rm D}-{\rm I})^{-1} \frac{1}{v_{n}} {\rm I}.
\nonumber
\end{eqnarray}
%
\subsection{Determine the unknown coefficients}
Compute all the terms in (\ref{definingrecursion}) to find the $c_i.$
Refer to \cite{WHetalcrm2004} for the details of the computation, 
resulting in
$ c_{2} = c_{5} = c_{6} = c_{7} = c_{8} = c_{10} = c_{11} 
= c_{12} = c_{13} = c_{15} = 0,$ and 
$ c_{1} = c_{3} = c_{4} = c_{9} = c_{14} = c_{16} = 1,$ and $c_{17} = -1.$
Substitute the constants into (\ref{rcandidatetoda}) to get 
(\ref{recursiontoda}).
%
\section{Conclusions and Future Research}
In this paper we presented algorithms for the symbolic computation of 
polynomial conservation laws, generalized symmetries, and recursion 
operators for systems of nonlinear DDEs.
We used the Toda lattice to illustrate the steps of the algorithms. 
The algorithms have been implemented in {\em Mathematica} and can be used 
to test the complete integrability of nonlinear DDEs.

Although our algorithm successfully finds conservation laws, generalized 
symmetries, and recursion operators for various Volterra and Toda lattices 
as well as the Ablowitz-Ladik lattice, the current recursion operator 
algorithm fails on nonlinear DDEs due to Belov and Chaltikian and 
Blaszak and Marciniak. 
In future research we intend to generalize the recursion operator algorithm 
so that it can cover a broader class of lattices.

\begin{acknowledgement}
J.A.\ Sanders, J.-P.\ Wang, M.\ Hickman, and B.\ Deconinck are gratefully
acknowledged for valuable discussions.
\end{acknowledgement}


%
%
%
%
%
%

%

\end{document}